\documentclass[11pt,letterpaper,usenames,dvipsnames]{article}
\usepackage[top=0.85in,left=0.75in,footskip=0.75in]{geometry}

\usepackage{amsmath,amssymb}

\usepackage{subfig,changepage}
\usepackage[utf8x]{inputenc}
 \usepackage{mathtools}
 \usepackage{nccmath}
\usepackage{textcomp,marvosym}
\usepackage{float} 
\usepackage{cite}

\usepackage{nameref,hyperref}

\usepackage[right]{lineno}
\usepackage{array}
\usepackage{afterpage}
\newcommand{\head}[1]{\textbf{#1}}
\newcolumntype{C}[1]{>{\centering\arraybackslash}p{#1}}

\usepackage[table]{xcolor}

\usepackage{array}

\newcolumntype{+}{!{\vrule width 2pt}}

\newlength\savedwidth

\raggedright
\usepackage[aboveskip=1pt,labelfont=bf,labelsep=period,justification=raggedright,singlelinecheck=off]{caption}

\makeatletter
\renewcommand{\@biblabel}[1]{\quad#1.}
\makeatother

\usepackage{lastpage,fancyhdr,graphicx}
\usepackage{epstopdf}
\fancyheadoffset[L]{2.25in}
\fancyfootoffset[L]{2.25in}
\begin{document}
\vspace*{0.2in}

\begin{flushleft}
{\Large
\textbf\newline{Topology of the mesoscale connectome of the mouse brain} 
}
\newline
\\
Pascal Grange
\\
\bigskip
Xi'an Jiaotong-Liverpool University, Department of Mathematical Sciences, Suzhou, China
\\
\bigskip

Pascal.Grange@xjtlu.edu.cn

\end{flushleft}

\section*{Abstract}
 The wiring diagram of the mouse brain 
 has recently been mapped at a mesoscopic scale in the Allen Mouse Brain Connectivity Atlas. Axonal projections
 from brain regions were traced using green fluoresent proteins. The resulting data were registered
  to a common three-dimensional reference space. They yielded  a matrix of connection strengths between 213 brain regions. 
  Global features such as closed loops formed by connections of similar intensity can be inferred using tools from  persistent homology.
  In this paper  the wiring diagram of the mouse brain is mapped  to a simplicial complex (filtered by connection strength),
 and  generators of the  first homology group are computed.
 Some regions, including nucleus accumbens, are connected to the entire brain by loops. 
   Thousands of loops go through the isocortex, the striatum and the thalamus.
 On the other hand, medulla is the only major  brain compartment that 
 contains more than 100 loops.



\section{Introduction}

The Allen Mouse Brain Connectivity Atlas \cite{AllenConnectome} has  filled a major
 gap in the knowledge of neuroanatomy by providing a brain-wide
 map of the connectome of the mouse brain. Adeno-associated 
 viral vectors expressing enhanced  green fluorescent protein were injected into the mouse brain, allowing to trace axonal projections.
  The scale of these experiments is mesoscopic \cite{positionPaper}, in the sense that injections target groups of neurons 
 belonging to brain regions assumed to be homogeneous. The microscopic
 scale would correspond to mapping individual synapses where neurons make contact. 
 The resulting three-dimensional fluorescent traces were registered to brain regions
  defined in the hierarchical Allen Reference Atlas \cite{ARA}.\\

 These results allowed the construction of the first  inter-region connectivity model of the mouse brain. 
 It takes the form of a   connectivity matrix, 
 whose rows and columns correspond to a brain region, and whose entries model the connection strengths
 between pairs of regions defined by classical neuroanatomy. The derivation 
 of the entries of of the connectivity matrix assumed homogeneity of brain regions  and additivity  of connection strengths.  
Projection densities correspond to axons
  highlighted by tracers, hence the projection densities
 from several different sources sum to produce projection density in a given region. This additivity assumption was used to address  experimental data
 in which an injection of viral tracer infected several neighbouring brain regions.\\

The construction 
  of the model from injection data (possibly infecting several regions for each injection) assumes that projections
 are homogeneous in each region, and that projections from different sources add  to produce projection density 
  to a given region. The entries of the connectivity matrix were observed to span a $10^5$-fold range, and are approximately 
 fit by a log-normal distribution.\\ 

    The connectivity matrix is naturally mapped to a graph that models the wiring diagram of the mouse brain. 
 Some features of the wiring diagram  
 such as node degree and clustering coefficient have been observed to be reproduced by scale-free networks \cite{scaleFree} and 
 small-world networks \cite{smallWorld}, but none of these models was found to fit all observed features of the wiring diagram.
 This begs for further modelling taking into account the global properties of the 
 connectivity map, together with the large range of intensities mapped by connection strengths. \\

  On the other hand, the recent computational  developments of algebraic topology \cite{Edelsbrunner1,Edelsbrunner2,computingPersistent,TDA} have given rise 
 to spectacular applications to data analysis in biological sciences, including 
 obstructions to phylogeny in viral evolution \cite{Rabadan}, and brain networks and neural correlations \cite{GiustiHippo,scaffolds,Giusti,GiustiHuman,BrainTopo}. 
In this paper we will apply techniques from algebraic topology, a branch of mathematics characterizing 
 properties of spaces that are invariant by continuous deformation, to work out 
 global properties of the wiring diagram of the mouse brain.

  We will first review the presentation of the Allen Mouse Brain Connectivity Atlas in the form 
 of a matrix modelling connection strengths between brain regions. We will define a mapping from these
 data to a filtered simplicial complex and explain why the generators of the first homology group are in one-to-one
 correspondence with closed loops in the mesoscale connectome. The anatomy of these loops will then be analysed
 by grouping loops according  to the major brain compartments they intersect. Moreover, we will  
 compare the fraction of the brain reached from each region by loops, to the fraction reached by direct connections
 encoded in the connectivity matrix.

\section{Materials and methods}
\subsection{Connectivity strengths in the mouse brain, local and global properties}
 Each image series in the Allen Mouse Connectivity Atlas  corresponds to an injection 
 of tracer, followed by sectioning and and imaging of the brain\footnote{
© 2014 Allen Institute for Brain Science. Allen Connectivity Atlas. 
Available from:  \href{http://connectivity.brain-map.org/}{\ttfamily{http://connectivity.brain-map.org/.org}}}.
 In \cite{AllenConnectome}, registered data from 469 image series
 were combined in order to estimate an inter-region connectivity matrix, presented in matrix form\footnote{The entries of the matrix $C$ correspond to the left-hand side of  Fig. 3 in \cite{AllenConnectome}.}:
\begin{equation}\label{connectivityMatrix}
C(r,r') = \left\{ {\mathrm{connection}}\;{\mathrm{strength}}\;{\mathrm{from}}\;  {\mathrm{region}}  \;{\mathrm{labelled}}
\; r\; {\mathrm{to}}\; {\mathrm{region}}
\; {\mathrm{labelled}}\; r'     \right\},\;\;\;\;1\leq r, r' \leq R.
\end{equation}
The size of this connectivity matrix is $R=213$, and each of the indices in $\{1,\dots,R\}$ corresponds  to a brain region,
 defined in the  
 Allen Reference Atlas (ARA, \cite{ARA,Swanson}).\\


The connectivity matrix captures the local structure of the connectome at mesocopic scale, 
 as it estimates the strength of direct connections from a region to other regions. To capture  global
  features of the connectome, we would like to identify closed circuits constructed from these connections.
  Algebraic topology formalises this notion: an irreducible  closed loop in a topological space (a loop that cannot be shrunk to a point 
 by continuous deformations) is a loop that is not the boundary of a disc in the topological space. It 
  is therefore a one-dimensional cycle that is not the boundary of a two-dimensional object. The family 
 of such objects in a topological space is invariant by continuous deformation, and has a group structure. It is called the 
   first homology group of the topological space, and denoted by $H_1$.\\

 If we repeat this reasoning in dimension zero, we obtain the more familiar notion of connected component: elements of  $H_0$ 
  are zero-dimensional objects that cannot 
  be joined by a  path drawn on the topological space.  They correspond to distinct connected components. 
 More generally, the elements  of the homology group $H_k$ are objects of dimension $k$ that are not boundaries of $(k+1)$-dimensional 
 objects in a topological space, and therefore formalise the notion of hole.
Moreover, the algebraic structure of  homology groups allows us to count independent objects in each of them: the number of generators of the  homology group
 $H_k$  is called the $k$-th Betti number and denoted by $b_k$. The Betti numbers $b_0$ and $b_1$ are the 
 number of connected components and the number of independent loops respectively.\\


\subsection{Mapping the matrix of connection strengths to a filtered simplicial complex}

  The connectivity matrix $C$ is not symmetric, because projections from a given brain region 
 to other brain regions are oriented (just as axons are). Let us  map  the connectivity matrix to a weighted graph with $R$ vertices
 corresponding to brain regions, and weighted edges corresponding to the entries of the connectivity matrix.  Edges 
 with two different orientations and different weights can exist between pairs of vertices.
  Let us  apply a decreasing function to the entries 
 of the connectivity matrix and define for instance:
\begin{equation}\label{distance}
d( r,s ) = -\log\left(  C( r,s) + \epsilon \right),
\end{equation}
 where $\epsilon$ is a positive regulator chosen to be smaller than the minimum connection strength, so that 
 entries of $d$ are bounded. The families of loops worked out in this paper do not depend on the choice of $\epsilon$  
 (in practice we took $\epsilon=0.01 {\mathrm{min}}_{r,s}C(r,s)$, resulting in a maximum entry of $42$ for $d$).
  This operation results in an approximately normal distribution of entries in the matrix $d$. 
The quantity $d(r,s)$ has been observed in \cite{AllenConnectome} to be positively correlated 
to the spatial distance between region labelled $r$  and region labelled $s$.
 However, the entries  of $d$ are not quite distances because  $d$ is not symmetric.\\


  We propose to map brain regions to a graph with $2R$ vertices as follows.
 Each  brain region, labelled $r$,  is mapped to a 
 pair of vertices $(s_r, t_r)$, where $s_r$ represents the sources of action potentials (the axons of the cell bodies in the region, that conduct axon potentials), and $t_r$
 the targets of action potentials (the dendrites in the region, that receive action potentials). 
 An edge  is drawn between vertices  $t_r$ and $s_r$ for each $r$ in $\{1,\dots,R\}$. With this doubling prescription we declare edges between targets and sources to be present in each region. We are going to look for closed paths in the wiring diagram of the brain, 
  constructed from the non-zero entries $C(s_r,t_r)$ of the connection matrix.
 Direct approaches to persistent 
 homology for asymmetric networks have been proposed, based on the Dowker complex, see \cite{Chowdhury}, but the present doubling prescription takes
 advantage of the mesoscopic scale of the connectome data.\\

 \begin{figure}[!ht]
\centering
     \subfloat[\label{subfig-1:dummy}]{%
       \includegraphics[width=0.8\textwidth]{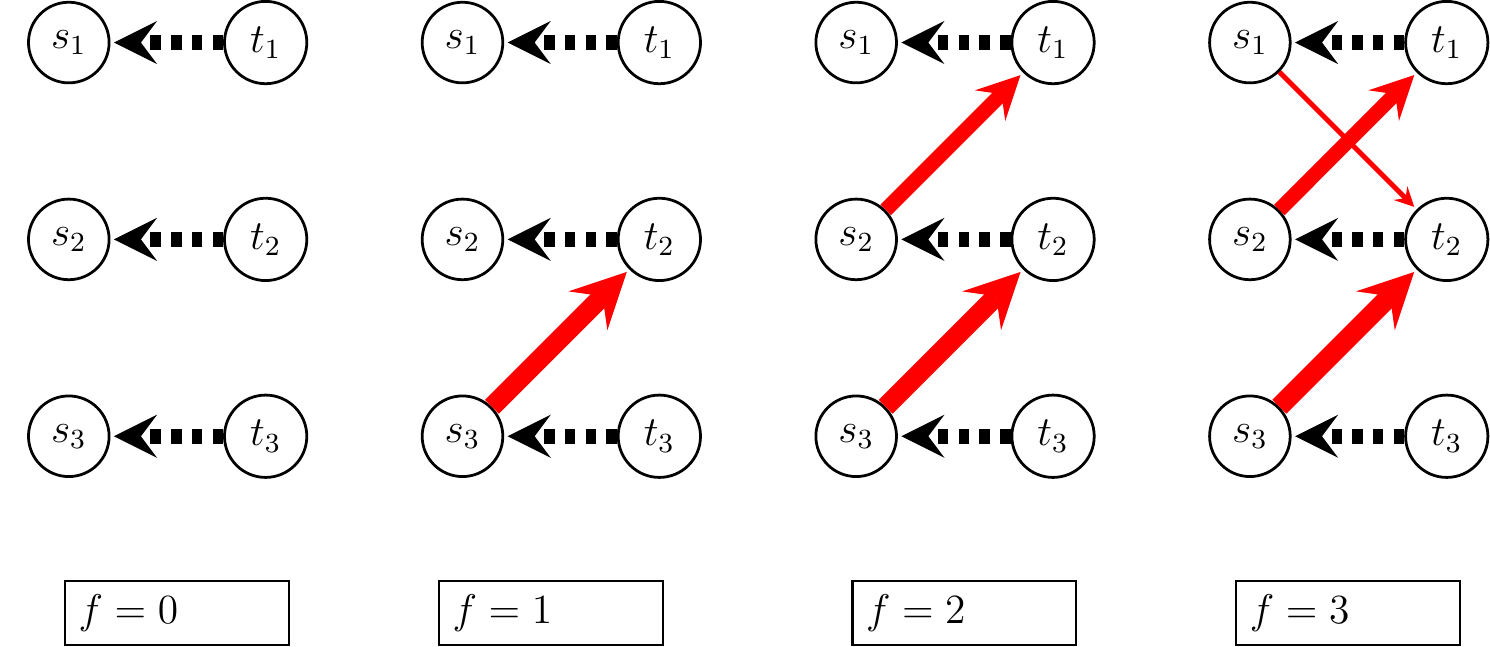}
     }
     \hfill
     \subfloat[\label{subfig-2:dummy}]{%
       \includegraphics[width=0.8\textwidth]{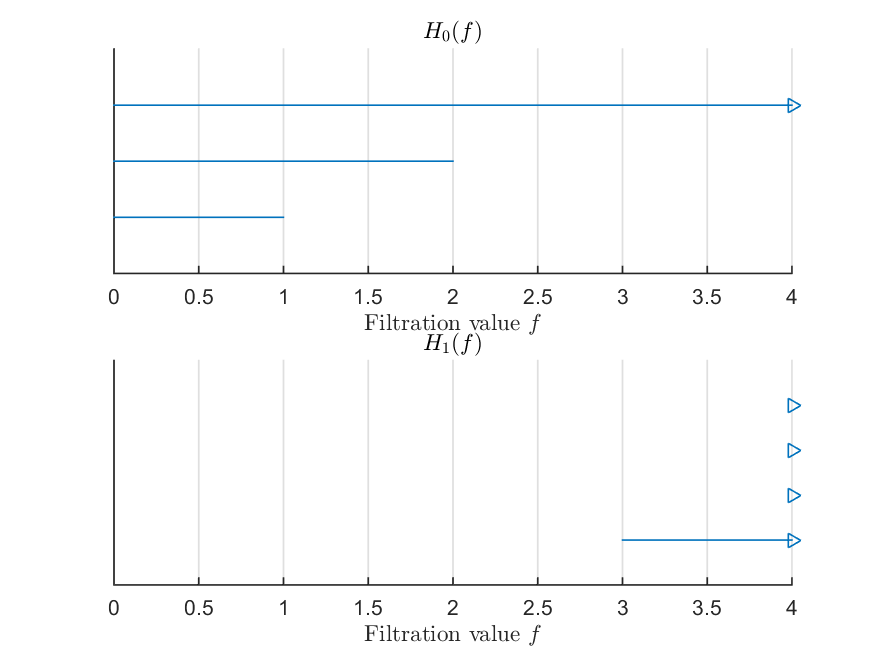}
     }
     \caption{{\bf{A toy model with three regions.}} (A) Each of the regions is doubled
 at filtration value zero, resulting in a source vertez and a target vertex. 
  The filtered complex start with three connected components
   and no loop.  For increasing using values of the filtration parameters, edges are drawn from to reflect projections from the nodes in the family $(s_r)_{1\leq r \leq 3}$ to nodes in
  the family $(s_r)_{1\leq r \leq 3}$. 
 (B) 
   The graph gets connected at filtration value $f=2$ (resulting in $b_0(f)=1$ for $f\geq 2$).
    A loop appears at $f=3$,  the corresponding generator of the first homology group is $[s_1,t_2] + [t_2,s_2 ] + [s_2, t_1 ] + [t_1,s_1] $. It goes through the 
 regions labelled $1$ and $2$.}
     \label{illustrateHomology}
   \end{figure}


   Given the family of $2R$ labelled vertices we have just described,  
 we can construct a family of graphs and work out the families of independent loops for each of them 
 by executing the following pseudo-code:\\
{\ttfamily{
1.  Consider a fixed $f \in ]0, {\mathrm{max}}_{1\leq r<s\leq R}d(r,s)]$ (referred to as a filtration value).\\
2. Draw an edge between any source labelled $s_a$ and any target labelled $t_b$ such that $d( s_a, t_b) \leq f$.\\
3. Work out generators of  the homology groups $H_0(f)$ (connected components) and $H_1(f)$  (independent loops) in the resulting graph.\\
}}
 The family of graphs (depending on the parameter $f$) is called a simplicial complex. 
 The third step of the procedure  uses techniques of simplicial homology, implemented
 in JavaPlex \cite{Javaplex,plexTutorial}.  Simplices in our case consist of vertices (zero-simplices, corresponding to brain regions),
  and edges (one-simplices, corresponding to axons connecting two brain regions). 
The result of the above procedure depends on the value of the filtration value $f$.
 At $f= 0$  we have as many connected components as points, and no loop. The number of connected components
 decreases when $f$  increases. It reaches $1$ for some value of the filtration value. When $f$ is increased beyond this value, 
  edges may still be drawn, possibly changing the number of loops.\\

When $f$ reaches the maximum entry of the matrix $d$, the simplicial complex cannot be changed anymore by increasing the filtration value.
  If we repeat the above procedure for all values of the entries of the matrix $d$, arranged 
 in increasing order, we can work out for each feature $F$  (which can be a connected component or a loop) an interval of 
 filtration values $[ f_{min}(F),f_{max}(F)]$ for which the 
  feature $F$ is present. These intervals can be drawn for all existing features $F$ of a fixed dimension, yielding 
 graphs of barcode form, one per dimension of feature (see Fig. \ref{illustrateHomology}B).
  Features persistent over a longer interval are thought  less likely to be due to noise \cite{barCode}. In particular, we will restrict ourselves to 
   cycles that persist from their first appearance (at a given filtration value) to the maximum entry of 
 the matrix  $d$.  The results of the above procedure are sketched of Fig. \ref{illustrateHomology} for a toy model 
 consisting of three regions (and  connection strengths assumed to be mapped to $d = \begin{pmatrix} 0& 3 & 4 \\ 2 & 0 & 4 \\ 4 & 1  &0\end{pmatrix}$). The pseudo-code 
 is executed for each unique value of the entries of $d$, in ascending order.\\



\subsection{Neuroanatomy of cycles}

\subsubsection{Three-dimensional presentation of brain regions}
 To analyse the anatomy of persistent cycles, 
 we used the voxelised version of the Allen Reference Atlas at a spatial  
 resolution of 25 microns \footnote{ See download instructions and code snippets in MATLAB and Python on the Allen Brain Atlas data portal:\\
 \ttfamily{http://help.brain-map.org/display/mouseconnectivity/API\#API-DownloadAtlas3-DReferenceModels }}. 
   This three-dimensional grid contains numerical labels encoding neuroanatomy at the 
 finest level compatible with the resolution  (the voxelised atlas consists of $V_{tot}\simeq 3.1\times 10^7$  cubic voxels,
  with 677 distinct regions). These numerical labels are associated to the names of brain regions accroding to the hierarchical annotation 
 of the ARA \footnote{The hierarchical system of annotation containing names of brain regions and numerical ids is available online 
 at  \href{http://api.brain-map.org/api/v2/structure\_graph\_download/1.json}{\ttfamily{http://api.brain-map.org/api/v2/structure\_graph\_download/1.json}}}. 
  For each of the $R=213$ regions in the mesoscale model 
 of the connectome (Eq. \ref{connectivityMatrix}), we resolved the numerical label 
  and the corresponding voxels.
 If the region has descendants in the hierarchical annotation,
 we lump together the voxels corresponding to these descendants, by annotating them with the 
 numerical label of the region. After this step, the voxelised atlas contains  $361$  distinct regions. 
  The  regions included in the mesoscale model of the connectome therefore do not quite span the entire 
 brain. Let us denote by  $V_r$ the number of voxels in region labelled $r$:
\begin{equation} 
 V_r=\left|\left\{v,\; {\mathrm{voxel}}\;{\mathrm{labelled}}\;v\; \in \;  {\mathrm{region}}\;{\mathrm{labelled}}\;r\right\}\right|,\;\;\;\;1\leq r\leq R.
\end{equation}
  The  brain regions corresponding to the rows of the connectivity matrix  contain 
\begin{equation}
 V = \sum_{r=1}^R V_r \simeq 2.4\times 10^7 \;{\mathrm{voxels}},
\label{numVoxDef}
\end{equation}
 about 79 percent of the total volume $V_{tot}$ of the  brain.  In the rest of this paper we will disregard the brain 
 regions that are not  included in the matrix of connectivity strengths, and the volume $V$ will be referred to as the volume of the 
 brain.\\ 

   Grouping 
 brain regions at a coarse hierarchical level  (according to  the ARA \cite{ARA, Swanson}) yields the following  major brain 
 compartments which will sometimes be designated by acronyms for presentation purposes:
 Isocortex,  olfactory areas (OLF),  cortical subplate (CTXsp), striatum (STR), pallidum (PAL), thalamus (TH),
   hypothalamus (HY), midbrain (MB), medulla (MY),  cerebellar cortex (CBX). These major brain compartments will be referred to as {\emph{big regions}}.\\


\subsubsection{Fraction of the brain connected by loops to a given region}

 Given a brain region labelled $r$ and a filtration value $f$,
  let us denote by $\mathcal{C}_r(f)$ the family of generators 
of the first homology group at filtration value $f$ that go through the region $r$:
   \begin{equation}
   {\mathcal{C}_r(f)} = \left\{  c \in H_1(f), r  \in c  \right\},
   \end{equation}
 and by $\overline{\mathcal{C}_r(f)}$ the set of vertices through which the elements of ${\mathcal{C}_r(f)}$ go:
   \begin{equation}
   {\overline{\mathcal{C}_r(f)}} = \left\{   s\in \{1,\dots,R\}, \exists c \in H_1(f), s \in c \right\}.
   \end{equation}
  In the toy model of Fig. \ref{illustrateHomology} with three brain regions, the sets  $\mathcal{C}_1(2)$ and $\mathcal{C}_3(3)$
 are empty. The set $\mathcal{C}_1(3)$ consists of one loop, corresponding to ${\overline{\mathcal{C}_r(f)}} = \{1,2\}$.\\

   We can calculate  the volume of the regions connected to the region $r$  by  loops, as a fraction 
 of the total number of voxels in the brain:
\begin{equation}
 \phi_r(f)=  \frac{1}{V}\sum_{s \in  \overline{\mathcal{C}_r(f)}} V_s.
\label{volFraction}
\end{equation}
 where $V$ is the total number of voxels defined in Eq. \ref{numVoxDef}, and $V_s$ is the number of voxels in region labelled $s$.  
 As the volumes of brain regions are not all equal, we can define an alternative measure by the fraction of the total number of regions 
  connected to region labelled $r$ by loops:
\begin{equation}
 \nu_r(f) =  \frac{1}{R}\left| \overline{ \mathcal{C}_r(f)}\right|.
\label{countingFraction}
\end{equation}
 For a  region labelled $r$ that is connected by  loops to all 
   brain  regions,  for some  value of the filtration $f$, we will have  $\phi_r(f) =  \nu_r(f) = 1$.\\

  It is natural  to compare 
 these ratios to the analogous ratios obtained by taking into account only direct connections between pairs of brain regions, as encoded by the connectivity 
matrix $C$. Let us denote these quantities by  $\phi^{(c)}_r(f)$ for the fraction of volume
 and by  $\nu^{(c)}_r(f)$ for the fraction of the number of regions: 
 \begin{equation}
 \phi^{(c)}_r(f) =  \frac{1}{V}\sum_{s = 1}^R V_s {\mathrm{max}}\left(\mathbf{1}( C(r,s ) >  f ) , \mathbf{1}( C(s,r ) > f ) \right),
\label{volConn}
\end{equation}
\begin{equation}
 \nu^{(c)}_r(f) =  \frac{1}{R}\sum_{s =1}^R{\mathrm{max}}\left(\mathbf{1}( C(r,s ) > f ) , \mathbf{1}( C(s,r ) > f ) \right),
\label{countingConn}
\end{equation}
 where ${\mathbf{1}}$ denotes the indicator function. 
 By construction the four fractions defined in Eqs \ref{volFraction},\ref{countingFraction},\ref{volConn},\ref{countingConn}
 are growing functions of the filtration value $f$. 
 There can be regions in the family $\overline{\mathcal{C}_r(f)}$  that are 
 not directly  connected to region $r$ based on the inter-region connectivity matrix. 
  There can also be regions with projections to or from region $r$ that are not involved in any 
 closed cycle going through region $r$ at the given filtration value $f$.  There is 
  therefore no a priori solidarity between the quantities $\phi$ and $\nu$,
 based on loops, and their analogues $\phi^{(c)}$ and $\nu^{(c)}$,  based  on direct connections.


\subsection{Brain-wide maps of  loops going through a given brain region}

To  sum up the connections by loops between two given brain  regions, 
 we can calculate the sum of connection strengths at which cycles 
 connecting these two regions appear, weighted by the numbers of such cycles. As the 
 generators of the first homology groups are independent, this is consistent with the 
 additivity assumption of connection strengths between brain regions. 
 For two  regions labelled $r$ and $s$ we introduce
\begin{equation}\label{weightedRegionByRegion}
 \omega(r,s)= \sum_{f}  e^{-f} \mathbf{1}(  s \in \overline{{\mathcal{C}}_r(f)} ),
\end{equation}
 where the sum is over the distinct  filtration values (the distinct entries of the 
 matrix $d$ defined in Eq. \ref{distance}), and the exponential function comes 
 from Eq. \ref{distance} relating connection strengths to the distance used to 
  defined filtration values.\\

 Fixing the region label $r$ and allowing the region label $s$ to take all the 
 possible values in $\{1,\dots, R\}$, we can define a brain-wide map of connection strengths  to region labelled $r$,
 by defining for voxel labelled $v$:
\begin{equation}\label{weightedRegionForVisu}
 \mathcal{W}_r (v)= \sum_f  e^{-f} \mathbf{1}(  s(v) \in  \overline{{\mathcal{C}}_r(f)} ),
\end{equation}
 where $s(v)$ is the index (in $\{1,\dots,R\}$) of the brain region to which voxel labelled $v$ belongs.  
   The quantity $\mathcal{W}_r$ maps  a voxel to a real number, and can be visualised 
 as a heat map for instance.\\

 To estimate how localised a brain-wide map of connection strengths is, we can compute the  Kullback--Leibler divergence
   from the above-defined profile and the uniform brain-wide density. If we denote by 
 the label of the region to which voxel labelled $v$ belongs, this divergence is expressed as 
\begin{equation}\label{KL}
 KL( r ) = \frac{1}{V}\sum_{v=1}^V   W_r(v)  \log( W_r(v) ),
\end{equation}
 where $V$ is the total number of voxels defined in Eq. \ref{numVoxDef}, and we used the fact that $W_r/V$ is 
 a normalised density function. This divergence induces a ranking of brain regions.

\section{Results}
\subsection{Persistent cycles}

\begin{figure}[h!]
\includegraphics[width=0.99\textwidth]{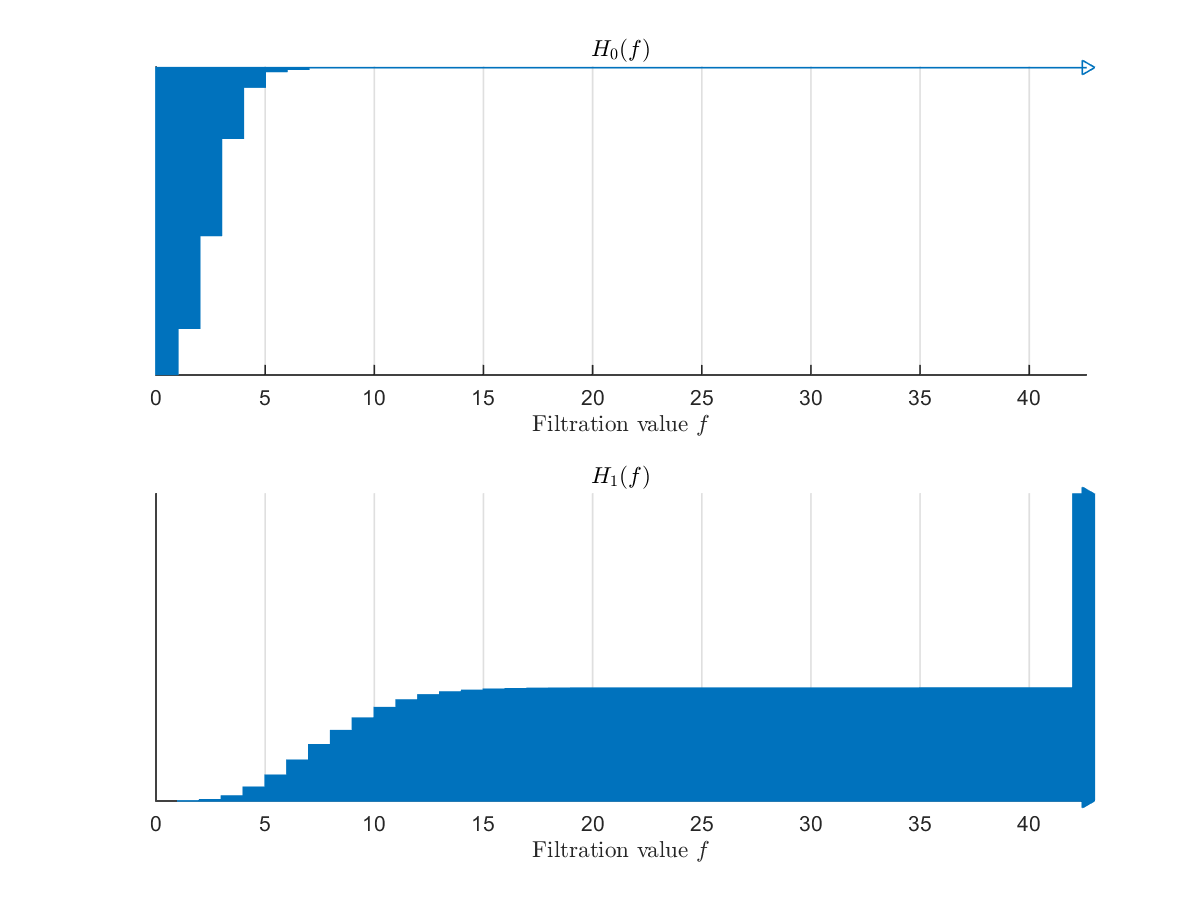}
\caption{{\bf{Bar code of the complex based on the inter-region connectivity matrix}}. The second plateau 
 that appears at filtration value $f=42$  corresponds to $-\log\epsilon$, where $\epsilon$ is the regulator introduced in Eq. \ref{distance}.
 The shape of the barcode and the families of generators of $H_1$ and $H_0$ would be unchanged if  $\epsilon$ went to zero.}
\label{figureFullCircuitBarCode}
\end{figure}

The bar code corresponding to the persistent generators of 
   the first two homology groups 
 is shown in Fig. \ref{figureFullCircuitBarCode}. The maximum Betti numbers
 are too large for the individual bars to be visible, as they are  on Fig. \ref{illustrateHomology}B.
  By construction the Betti number $b_0$ starts at $R=213$. The Betti number $b_1$ starts at $0$ 
 and the plateau at filtration value $25$
  in the bar code of the first homology group corresponds to $b_1=16,529$. At filtration values larger than
  $f_{conn} = 7$, all brain regions are in the same connected component.
 Moreover, from the growth profile of the bar code 
 of the first homology group, we can see that thousands of persistent cycles appear at filtration values lower than $f_{conn}$ 
 ($b_1=8,230$ at $f=f_{conn}$).

\subsection{Loops highlight the cortico-striato-thalamic network}
  To obtain a coarse picture of the  anatomy of cycles,
    let us  work out which family of {\emph{big regions}} (as defined in Section 2.3.1) is intersected by 
 each loop. We can rank the resulting families of big 
 regions by decreasing value of prevalence.  For an example of the brain regions in a loop appearing at filtration value $f=2$, 
 together with the names of the corresponding big regions, see Table \ref{tableForCircuit7}. This loop appears at a low filtration 
 value and we will see that the family of big regions it intersects (isocortex, striatum, thalamus and midbrain) is frequent among loops in the wiring diagram 
 of the mouse brain.\\ 

\begin{table}[!ht]
\centering
\caption{
{\bf Brain regions in a circuit appearing at filtration value $f=2$.}}
\begin{tabular}{|l|l|}
\hline
\textbf{Brain region}&\textbf{Big region}\\\hline
Anterior cingulate area, dorsal part&Isocortex\\\hline
Primary auditory area&Isocortex\\\hline
Caudoputamen&Striatum\\\hline
Medial geniculate complex, dorsal part&Thalamus\\\hline
Medial geniculate complex, ventral part&Thalamus\\\hline
Superior colliculus, motor related&Midbrain\\\hline
\end{tabular}
\\
\begin{flushleft}Six regions appear, belonging to  four distinct {\emph{big}} brain regions.
\end{flushleft}
\label{tableForCircuit7}
\end{table}

  For definiteness we did this calculation at two special filtration values:\\
-  the filtration value $f_{conn}=7$ at which the Betti number $b_0$ falls to 1.\\
- the maximum filtration value $(f_{max} = 37)$, above which no extra loops appear (before the value $-\log\epsilon = 42$ introduced in Eq. \ref{distance} to 
 provide a cut-off to the plots).\\
  We found that $334$ distinct  families of big regions occur, but the most frequent 27  families account for 
 50\% of the loops.  
 The most prevalent families of big regions are presented in Table  \ref{coarseAnalysis} (where they are ordered by decreasing 
 prevalence at filtration value $f=7$).  We notice that midbrain is represented in most of the rows of this table\footnote{Midbrain is also represented in most of the
 loops that appear at low filtration values $f<3$.}, and that  the three most frequent combinations (accounting for more than $10\%$ of all the loops) 
 go through six or more big regions.
 The family of four big regions corresponding to the regions in the loop presented in Table \ref{tableForCircuit7}
 is the fourth most prevalent family of big regions in loops appearing at filtration values lower than $f=7$, with 249 distinct loops  (resp. fifth, $f=37$ and 328 loops). 
  If  we focus on small 
 families of big regions (with three regions or fewer), we observe that the most prevalent is the cortico-striato-thalamic family (7th row of the table, with 183 loops at $f=7$).
  Highlighting occurrences of  isocortex, striatum and thalamus in colour in Table  \ref{tableForCircuit7}  (in red, green and blue respectively), we notice 
 that the first 17 rows present at least one of these colours (in 14 cases associated with midbrain, highlighted in brown). The next small family 
  consists of isocortex and striatum (16th rank, with 106 loops), followed at rank 18 by medulla. With 104 loops, medulla contains more cycles 
 than the cortico-thalamic family (ranked 25 with 80 loops).  Moreover, medulla is the only single big region to support 
 more than 10 cycles (see Table \ref{coarseAnatomy1}, which shows that only $147 $ cycles are confined to a single big region, out of which 134 appear before the filtration value 7).\\

\begin{table}[!ht]
\begin{adjustwidth}{-0.5in}{0in} 
\centering
\caption{
{\bf  Families of big regions containing loops.}}
\begin{tabular}{|C{0.8cm}|C{1.5cm}|C{1.9cm}|C{1.5cm}|C{1.9cm}|C{8.2cm}|}
\hline
\textbf{Rank (out of 334)}&\textbf{\head{Number of loops ($f=7$)}}&\textbf{\head{Cumulated percentage of loops ($f=7$)}}&\textbf{\head{Number of loops ($f=37$)}}&\textbf{\head{Cumulated percentage of loops ($f=37$)}}&\textbf{\head{Big regions intersected}}\\\hline
1&410&6\%&600&4\%&Cortical Subplate, Hippocampal Formation, Hypothalamus, {\textcolor{red}{Isocortex}}, {\textcolor{Bittersweet}{Midbrain}}, {\textcolor{ForestGreen}{Striatum}}, {\textcolor{Blue}{Thalamus}}\\\hline
2&268&9\%&614&8\%&Cortical Subplate, Hypothalamus, {\textcolor{red}{Isocortex}}, {\textcolor{Bittersweet}{Midbrain}}, {\textcolor{ForestGreen}{Striatum}}, {\textcolor{Blue}{Thalamus}}\\\hline
3&265&12\%&455&11\%&Cortical Subplate, Hypothalamus, {\textcolor{red}{Isocortex}}, {\textcolor{Bittersweet}{Midbrain}}, Pallidum, {\textcolor{ForestGreen}{Striatum}}, {\textcolor{Blue}{Thalamus}}\\\hline
4&249&15\%&328&13\%&{\textcolor{red}{Isocortex}}, {\textcolor{Bittersweet}{Midbrain}}, {\textcolor{ForestGreen}{Striatum}}, {\textcolor{Blue}{Thalamus}}\\\hline
5&248&19\%&346&15\%&Hypothalamus, {\textcolor{red}{Isocortex}}, {\textcolor{Bittersweet}{Midbrain}}, Pallidum, {\textcolor{ForestGreen}{Striatum}}, {\textcolor{Blue}{Thalamus}}\\\hline
6&199&21\%&283&17\%&Hypothalamus, {\textcolor{red}{Isocortex}}, {\textcolor{Bittersweet}{Midbrain}}, {\textcolor{ForestGreen}{Striatum}}, {\textcolor{Blue}{Thalamus}}\\\hline
7&183&23\%&222&19\%&{\textcolor{red}{Isocortex}}, {\textcolor{ForestGreen}{Striatum}}, {\textcolor{Blue}{Thalamus}}\\\hline
8&165&25\%&255&20\%&Cortical Subplate, {\textcolor{red}{Isocortex}}, {\textcolor{Bittersweet}{Midbrain}}, {\textcolor{ForestGreen}{Striatum}}, {\textcolor{Blue}{Thalamus}}\\\hline
9&141&27\%&175&21\%&Hypothalamus, {\textcolor{red}{Isocortex}}, {\textcolor{Bittersweet}{Midbrain}}, {\textcolor{Blue}{Thalamus}}\\\hline
10&135&29\%&175&22\%&Hypothalamus, {\textcolor{Bittersweet}{Midbrain}}, Pallidum, {\textcolor{Blue}{Thalamus}}\\\hline
11&130&31\%&190&24\%&Cortical Subplate, Hippocampal Formation, Hypothalamus, {\textcolor{Bittersweet}{Midbrain}}, {\textcolor{ForestGreen}{Striatum}}, {\textcolor{Blue}{Thalamus}}\\\hline
12&130&32\%&208&25\%&Cortical Subplate, Hippocampal Formation, Hypothalamus, {\textcolor{red}{Isocortex}}, {\textcolor{Bittersweet}{Midbrain}}, Pallidum, {\textcolor{ForestGreen}{Striatum}}, {\textcolor{Blue}{Thalamus}}\\\hline
13&122&34\%&149&26\%&Cortical Subplate, {\textcolor{red}{Isocortex}}, {\textcolor{ForestGreen}{Striatum}}\\\hline
14&115&35\%&163&27\%&Hypothalamus, {\textcolor{red}{Isocortex}}, {\textcolor{Bittersweet}{Midbrain}}, Pallidum, {\textcolor{Blue}{Thalamus}}\\\hline
15&107&37\%&135&28\%&Cortical Subplate, Hypothalamus, {\textcolor{red}{Isocortex}}, {\textcolor{Bittersweet}{Midbrain}}, Pallidum, {\textcolor{ForestGreen}{Striatum}}\\\hline
16&106&38\%&117&29\%&{\textcolor{red}{Isocortex}}, {\textcolor{ForestGreen}{Striatum}}\\\hline
17&105&39\%&139&29\%&Cortical Subplate, Hippocampal Formation, Hypothalamus, {\textcolor{Bittersweet}{Midbrain}}, Olfactory Areas, {\textcolor{ForestGreen}{Striatum}}, {\textcolor{Blue}{Thalamus}}\\\hline
18&104&40\%&115&30\%&{\textcolor{Magenta}{\large\bf{Medulla}}}\\\hline
19&96&42\%&108&31\%&Hypothalamus, {\textcolor{Bittersweet}{Midbrain}}, {\textcolor{Blue}{Thalamus}}\\\hline
20&85&43\%&104&32\%&Hypothalamus, {\textcolor{Bittersweet}{Midbrain}}, Pallidum\\\hline
21&85&44\%&111&32\%&Hypothalamus, {\textcolor{red}{Isocortex}}, {\textcolor{Bittersweet}{Midbrain}}, Pallidum, {\textcolor{ForestGreen}{Striatum}}\\\hline
22&84&45\%&118&33\%&{\textcolor{Magenta}{\large\bf{Medulla}}}, {\textcolor{Bittersweet}{Midbrain}}, Pallidum\\\hline
\end{tabular}
\\
\begin{flushleft} The families of big regions are  ordered by decreasing number of loops they contain (at filtration value $f=7$).
\end{flushleft}
\label{coarseAnalysis}
\end{adjustwidth}
\end{table}

\begin{figure}[!ht]
\includegraphics[width=0.95\textwidth]{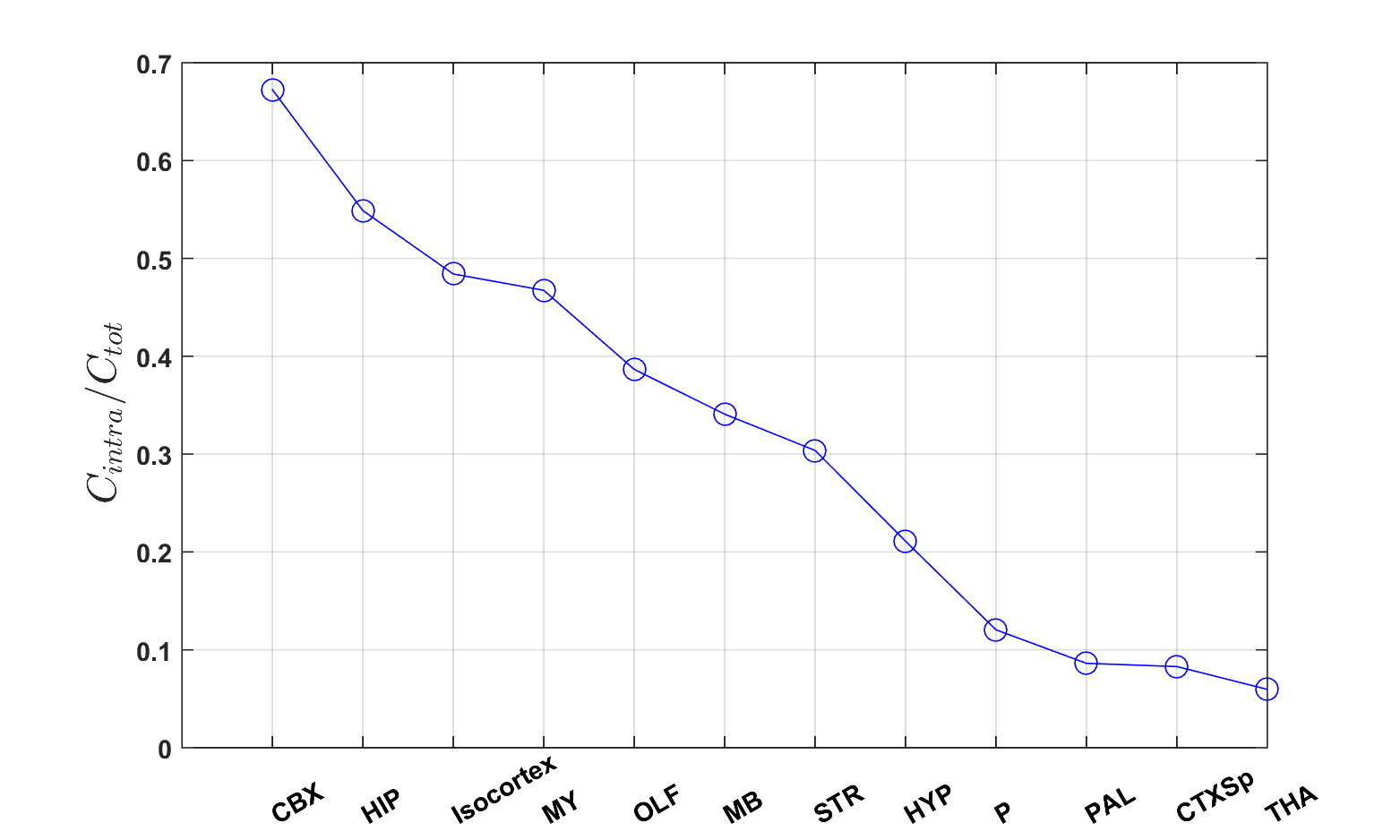}
\caption{Fraction of inter-region connection strengths supported in the big region of origin. Acronyms of big regions read as follows: OLF = olfactory areas,
 HPF = hippocampal formation, CTXsp = cortical subplate, STR = striatum, PAL = pallidum, TH = thalamus, HY = hypothalamus, 
 MB = midbrain, MY = medulla, CBX = cerebellar cortex.}
\label{confinedConnections}
\end{figure}

  To assess whether the dominance of medullar loops among loops confined to a single big region 
  could have been guessed based on the matrix of connectivity strengths,
  let us calculate how much 
 of the connection strengths from its sub-regions project within the same region. For each big region labelled $b$, let us define:
\begin{equation}
\frac{{\mathcal{C}}_{intra}}{{\mathcal{C}}_{tot}}(b) = \frac{\sum_{r=1}^R\sum_{s=1}^R{\mathbf{1}}( B(r) = b ){\mathbf{1}}( B(s) = b )  C(r,s)}{\sum_{r=1}^R\sum_{s=1}^R {\mathbf{1}}( B(r) = b ) C(r,s)},
\end{equation}
  where the symbol ${\mathbf{1}}( B(r) = b)$ equals $1$ if the region labelled $r$ is part of the big region labelled $b$ (for instance we can read from the third row of Table \ref{tableForCircuit7} that ${\mathbf{1}}( B(\mathrm{Caudoputamen}) = \mathrm{Striatum}) = 1$). 
  The ratio ${\mathcal{C}}_{intra}/{\mathcal{C}}_{tot}$ is the conditional probability of a connection  from within the big region $b$ 
  to project within region $b$.  
  The sorted values of this ratio are
 plotted on Fig. \ref{confinedConnections}. The average value is $31\%$,  and the values 
  range from $6\%$  (thalamus) to $68\%$ (cerebellar cortex, which is found to contain only one loop, see Table \ref{coarseAnatomy1}B).\\
 The fact that 80 percent of the cycles  confined to a single big region are in medulla 
  could not have been guessed based on the connectivity matrix only: $47\%$ of connections 
  from medullar regions project to medullar regions, which is above average, but the medulla is only 
  ranked fourth among 12 big regions by our measure of conditional probability.\\
  
%
%
%
%
%



%
\begin{table}[!ht]
\begin{adjustwidth}{2.25in}{0in} 
\caption{
{\bf Tables of single big regions containing loops, ordered by decreasing number of loops.}}
\begin{tabular}{|l|l|l|}
\hline
\textbf{Number of loops}&\textbf{Fraction (\%)}&\textbf{Big region label}\\\hline
104&78&Medulla\\\hline
8&6&Hippocampal Formation\\\hline
8&6&Hypothalamus\\\hline
6&5&Olfactory Areas\\\hline
5&4&Isocortex\\\hline
3&3&Midbrain\\\hline
\end{tabular}
\\
(A)\\
\begin{tabular}{|l|l|l|}
\hline
\textbf{Number of loops}&\textbf{Fraction (\%)}&\textbf{Big region label}\\\hline
115&79&Medulla\\\hline
9&7&Hypothalamus\\\hline
8&6&Hippocampal Formation\\\hline
6&5&Olfactory Areas\\\hline
5&4&Isocortex\\\hline
3&3&Midbrain\\\hline
1&1&Cerebellar Cortex\\\hline
\end{tabular}
\\
(B)\\
\begin{flushleft}  (A) For filtration value $f=7$, there is a total of 134 cycles. (B) For filtration value 
 $f=37$, there are 147 loops, and just one more region represented.
\end{flushleft}
\label{coarseAnatomy1}
\end{adjustwidth}
\end{table}

\subsection{Loops can connect brain regions to the entire brain}

\begin{figure}[h!]
\includegraphics[width=0.9\textwidth]{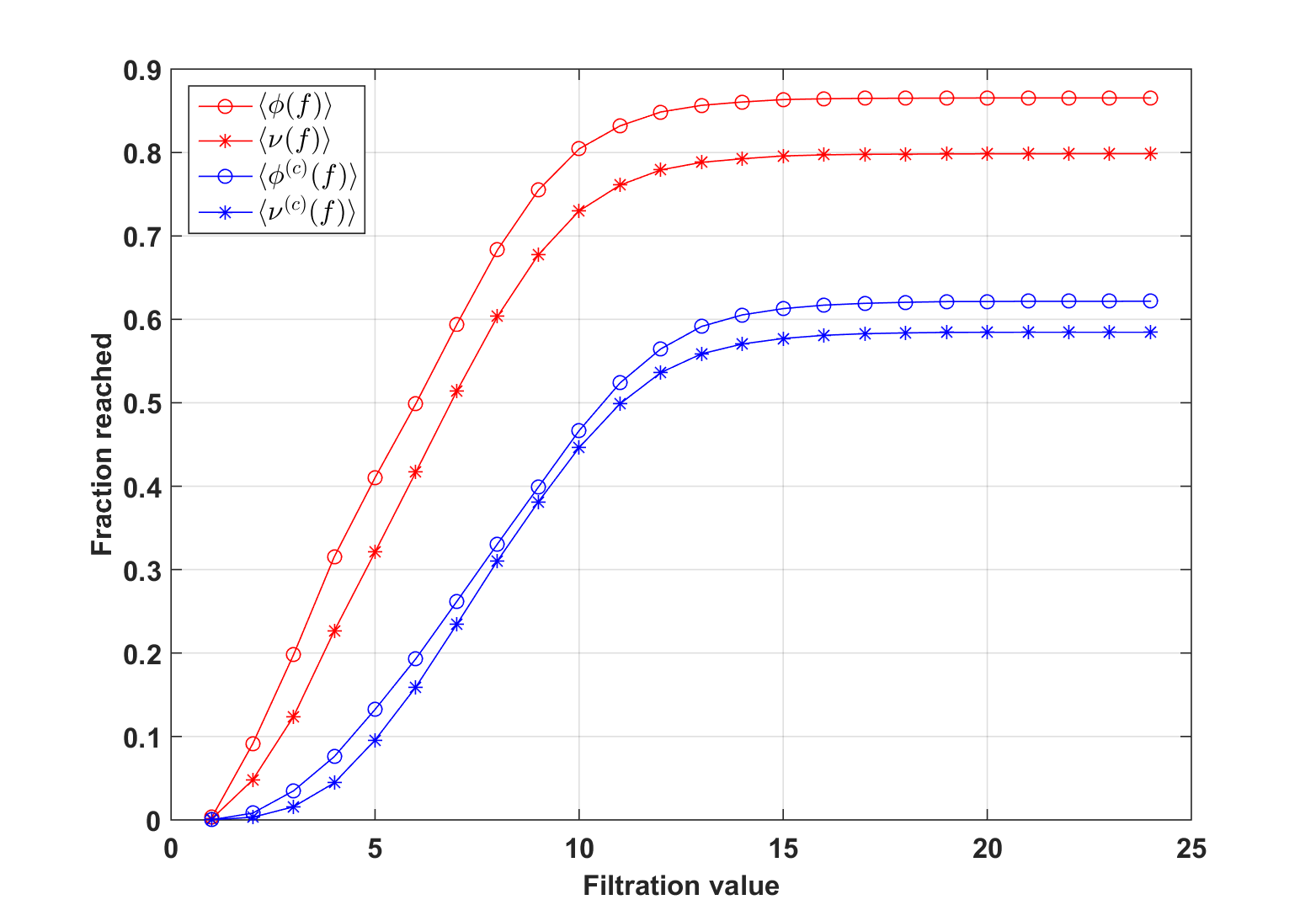}
\caption{Fraction of the brain reached by loops, averaged across all regions in the connectivity 
 atlas.}
\label{figVolAverageDirected}
\end{figure}

 On average,  loops connect a given brain region to a larger domain in the brain 
  than direct projections.
This is true for both the volumetric and the counting measures
 defined in Eqs \ref{volConn} and \ref{countingConn},
 and at all filtration values.
  This can be observed on Fig. \ref{figVolAverageDirected}, where the following averages 
 across all brain regions are plotted:
\begin{equation}
\begin{split}
 \langle \phi(f)\rangle:&= \frac{1}{R}\sum_{i=1}^R\phi_r(f), \;\;\; \langle \nu(f)\rangle = \frac{1}{R}\sum_{i=1}^R\nu_r(f),\\
\langle \phi^{(c)}(f)\rangle&= \frac{1}{R}\sum_{i=1}^R\phi^{(c)}_r(f), \;\;\; \langle \nu^{(c)} (f) \rangle = \frac{1}{R}\sum_{i=1}^R\nu^{(c)}_r(f).
\end{split}
\end{equation}
  Moreover, all four averages reach an asymptote, and the ratios of asymptotic values do not depend heavily
 on the choice of measure, as:\\
\begin{equation}
\langle \phi (25)\rangle  \simeq 1.39 \langle \phi^{(c)}(25) \rangle\simeq 87\%,\;\;\;\; {\mathrm{and}}\;\;\;\; \langle \nu(25) \rangle \simeq 1.37 \langle \nu^{(c)}(25)\rangle \simeq 80\%.
 \end{equation}

\begin{figure}[h!]
\includegraphics[width=0.95\textwidth]{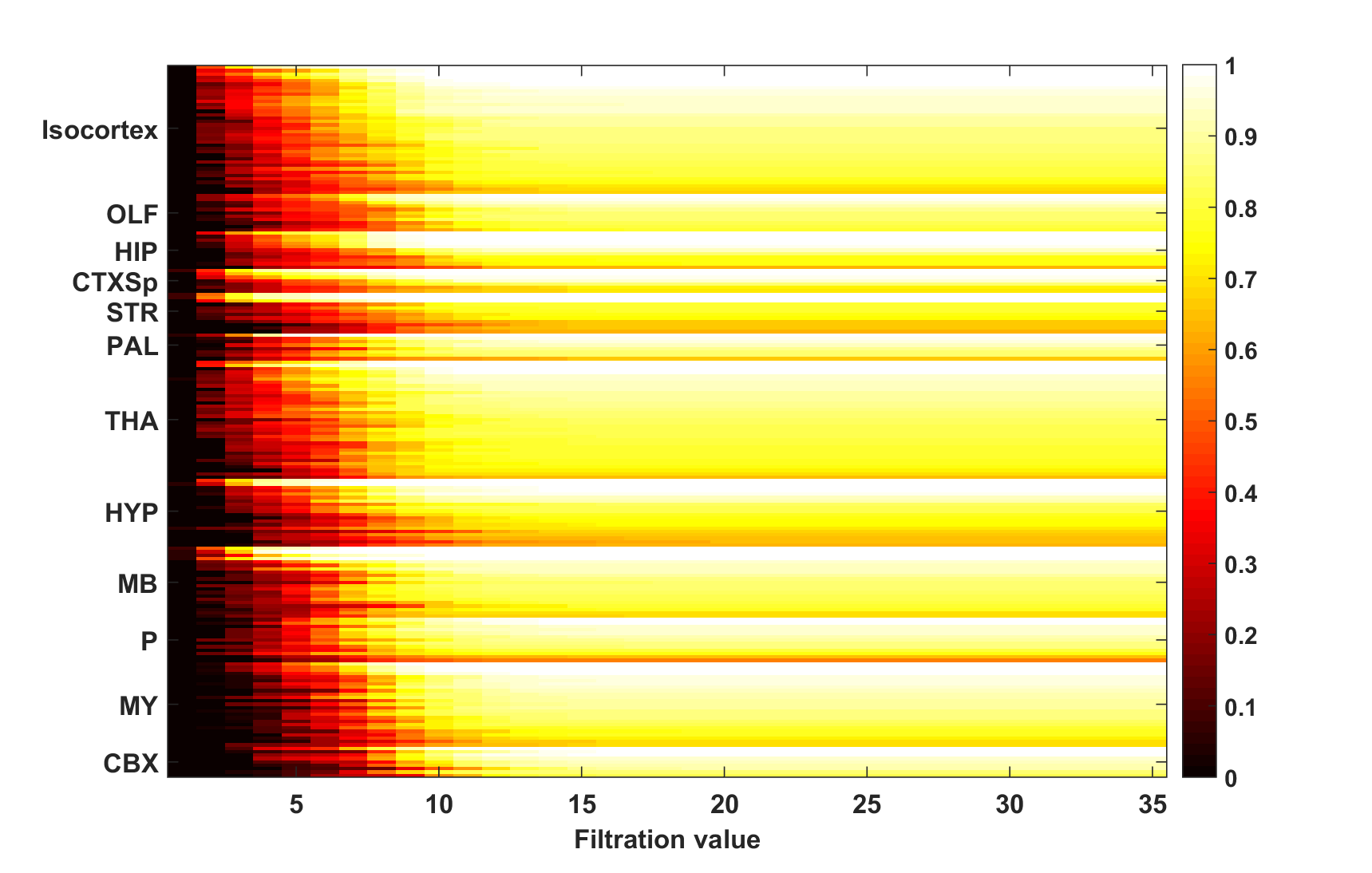}
\caption{{\bf{Heat maps of the fraction of the brain reached by loops $\phi_r(f)$, as a function of the filtration value (horizontal axis).}} The rows 
 correspond to brain regions (grouped by big regions according to coarse neuroanatomy). The subregions in each big region are ordered by decreasing value 
  of the asymptote  $\phi_r(35)$ .}
\label{figVolHeatMapDirected}
\end{figure}

However, the maximum values of the fractions $\phi_r$ and $\nu_r$ are highly heterogeneous among brain regions. 
 The heat map of Fig. \ref{figVolHeatMapDirected} shows the 
 fraction of the brain $\phi_r(f)$, with the region labels $r$ grouped 
 by big regions (and ordered within each big region by decreasing value of the maximum reached at large filtration value).
In particular, 18 regions reach the entire brain through connections by loops. 
 Their names are shown on Table \ref{tableMaxVolume}, together with the big regions to which they belong (midbrain is the most represented big region, with 4 regions).
  The ansiform lobule, which is part of the cerebellar cortex, is connected to the entire brain by loops, and is the brain region connected to the 
 largest volume in the brain based on the connectivity matrix (it is ranked first by  the measure $\phi^{(c)}$). However, not all the regions appearing
 in Table \ref{tableMaxVolume} are that strongly connected based on the connectivity matrix. For instance the posterior amygdalar nucleus is ranked 
 173 out of 213 by the measure $\psi^{(c)}$).\\

\afterpage{\clearpage}
\begin{table}[!ht]
\begin{adjustwidth}{-0.5in}{0in} 
\centering
\caption{
{\bf Table of brain regions connected to all brain regions by loops.}}
\centering 
\begin{tabular}{|l|l|l|l|}
\hline
\textbf{Big region}&\textbf{Name of region}&\textbf{$\phi^{(c)}_r(25) (\%)$}&\textbf{Rank by $\phi^{(c)}$, out of $213$}\\\hline
Isocortex&Anterior cingulate area, dorsal part&85&13\\\hline
Hippocampal Formation&Entorhinal area, lateral part&81&31\\\hline
Cortical Subplate&Basolateral amygdalar nucleus&81&29\\\hline
Cortical Subplate&Posterior amygdalar nucleus&50&173\\\hline
Striatum&Nucleus accumbens&68&73\\\hline
Striatum&Caudoputamen&55&148\\\hline
Striatum&Medial amygdalar nucleus&59&123\\\hline
Pallidum&Globus pallidus, internal segment&77&40\\\hline
Thalamus&Medial geniculate complex, dorsal part&71&64\\\hline
Thalamus&Peripeduncular nucleus&74&52\\\hline
Hypothalamus&Lateral hypothalamic area&70&67\\\hline
Hypothalamus&Subthalamic nucleus&73&56\\\hline
Midbrain&Central linear nucleus raphe&75&49\\\hline
Midbrain&Midbrain reticular nucleus&56&138\\\hline
Midbrain&Periaqueductal gray&66&87\\\hline
Midbrain&Superior colliculus, motor related&59&125\\\hline
Cerebellar Cortex&Ansiform lobule&98&1\\\hline
\end{tabular}

\begin{flushleft}  On average these brain regions are connected to $72\%$ of the volume of the brain
 by direct connections, which is largest than the average value $\langle \phi^{(c)}\rangle\simeq 63\%$. However, 5 regions correspond to values below this average.
\end{flushleft}
\label{tableMaxVolume}
\end{adjustwidth}
\end{table}

\begin{figure}[h!]
\centering
     \subfloat[\label{subfig-1:dummy}]{
       \includegraphics[width=0.8\textwidth]{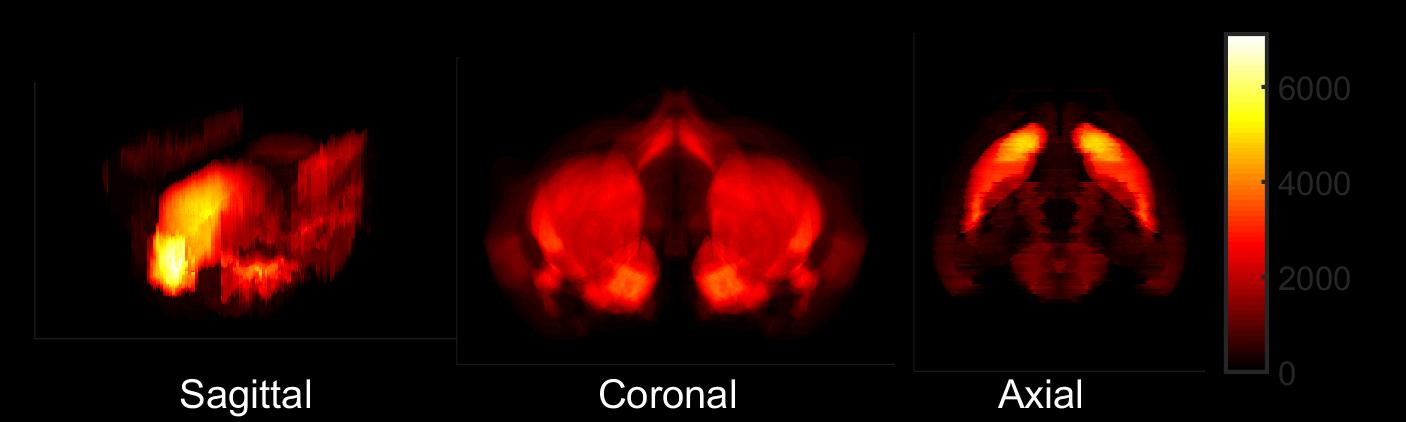}
     }
     \hfill
     \subfloat[\label{subfig-1:dummy}]{
       \includegraphics[width=0.8\textwidth]{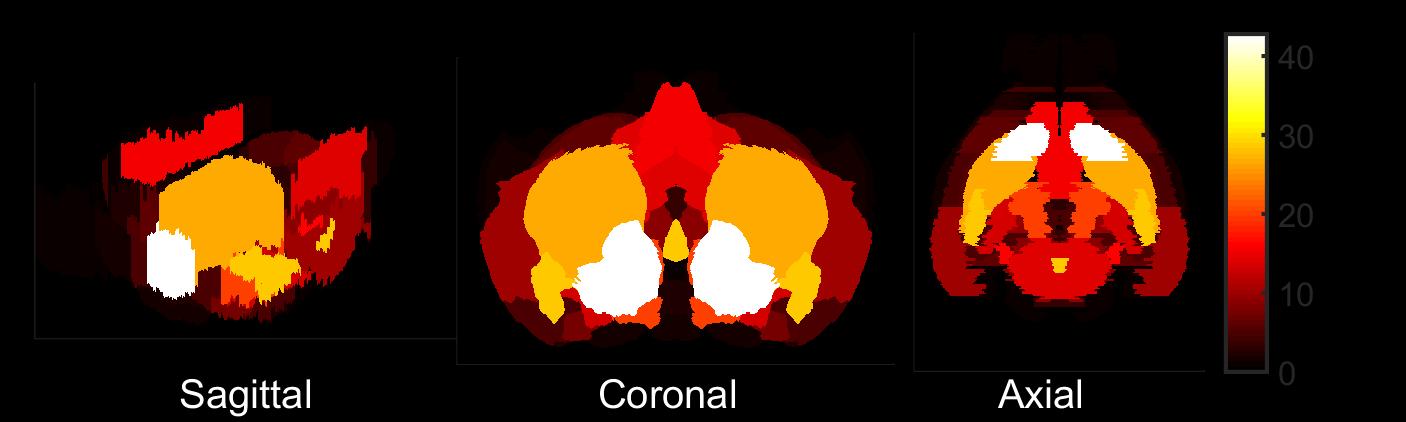}
     }
     \hfil
     \subfloat[\label{subfig-2:dummy}]{
       \includegraphics[width=0.8\textwidth]{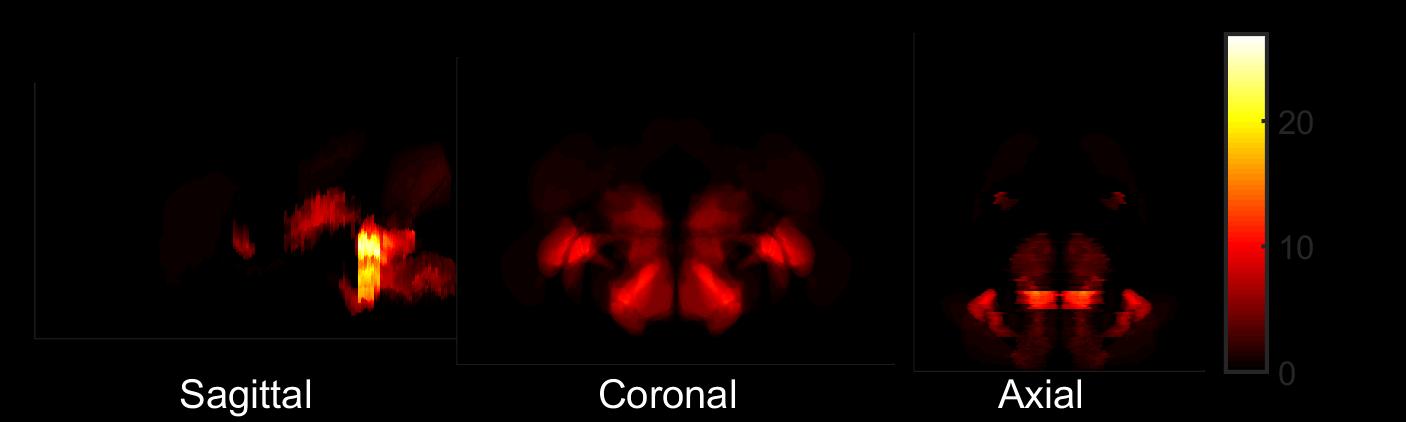}
     }
     \hfill
     \subfloat[\label{subfig-2:dummy}]{
       \includegraphics[width=0.8\textwidth]{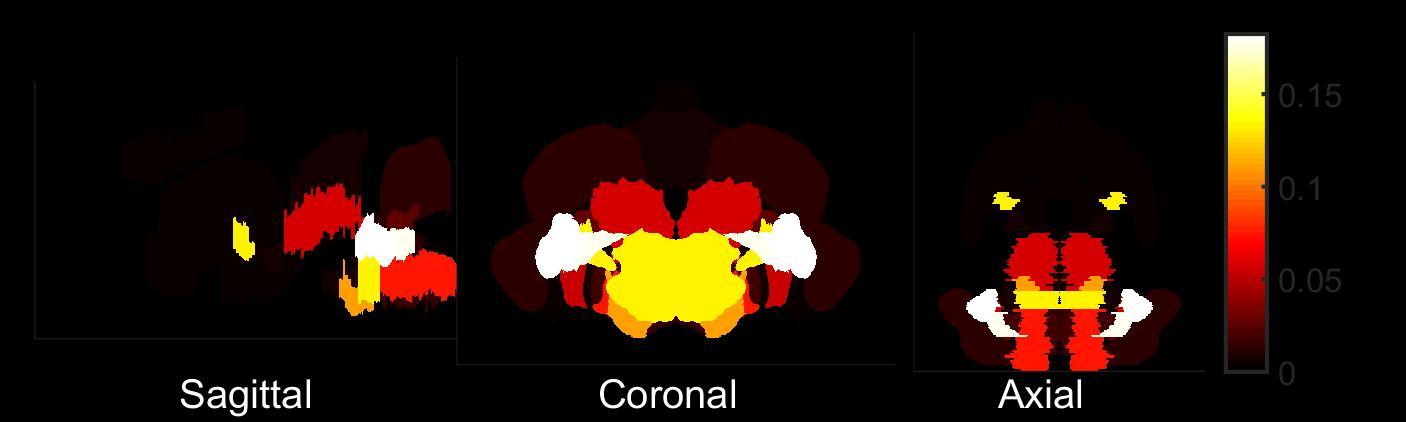}
     }
     \caption{{\bf{Heat map of the weighted regions connected by loops to given brain regions, projected in the sagittal, coronal and axial directions. }}  (A) Nucleus accumbens, 
    projection of sums of intensities. (B) Nucleus accumbens,  projection of maximal intensities. (C) Flocculus, projection of sums  of intensities. (D) Flocculus,  projection of maximal intensities.
        For visualisation purposes, both hemispheres are shown, even though the 
  results are based on the ipsilateral connectivity matrix only.}
     \label{illustrateBrainwide}
   \end{figure}
\afterpage{\clearpage}

\subsection{Brain-wide density of connections by loops to a given region}
 The region for which $W_r$ has lowest Kullback--Leibler divergence from a uniform profile 
   is the nucleus accumbens. The region with  highest Kullback--Leibler divergence from a uniform profile is the flocculus, 
 a region in the cerebellar cortex. 
  Heat maps of the densities of loops for both regions are shown on Fig. \ref{illustrateBrainwide}. They 
 are projected on sagittal, coronal and axial planes. 
 For each of these regions the volume can be presented either as the sum of 
 values of voxels (Figs \ref{illustrateBrainwide}A,C) or as maximal-intensity projections (Figs \ref{illustrateBrainwide}B,D).
  On maximal-intensity projections of the density $W_r$, the voxels belonging to region labelled $r$ 
  have the largest value by construction (because all loops go through this region). Maximal-intensity projections 
 can exhibit at most $R$ different values (they are piecewise constant on each brain region).
 On the other hand, projections of sums of the density $W_r$  may have  higher values in voxels that do not 
 belong to region labelled $r$, depending on the direction of projection. Even though nucleus accumbens is connected
 to the entire brain by loops (meaning all the voxels in Figs. \ref{illustrateBrainwide}A,B correspond to non-zero values),
 the plots still appear rather shallow, and the strength of connection by loops is still far from uniform, 
 even  in the region that minimises the divergence from a uniform brain-wide profile.\\

\section{Discussion}




 Persistent homology reveals global properties of the mouse connectome at the mesoscopic scale.
   The high prevalence of cortico-striato-thalamic loops is a strong consistency check between the 
  mesoscale experimental approach  \cite{positionPaper,AllenConnectome} and the present topological method,
 as the cortico-striato-thalamo-cortical circuit 
  is known as a major circuit regulating  complex 
  behaviours (for a review of its involvement in the Tourette sydrome see \cite{Tourette}).\\

 The global nature of topological tools reveals connections  between 
 regions  that are not linked by direct projections. Moreover,  
  the lists of generators of the first homology groups allow to count the 
   number of independent loops for a given threshold in connection strengths. 
   The relative abundance of  closed loops in the medulla is perhaps 
 best interpreted in terms of the life-sustaining automatic  functions regulated by the medulla.
  The corresponding circuits are topologically insulated from circuits involved in conscious behaviours or 
 learning \cite{pontoMedullar}.\\



  Axons, in addition to conducting action potentials,  transport biological molecules \cite{whiteMatterTransport,Huang}. 
Integration with other brain-wide data sets, in particular gene-expression data \cite{AllenGenome,AllenAtlasMol}, has already 
 shown that connected regions in the adult rodent brain have increased similarity in gene-expression profiles, 
 and sets of genes most correlated to connectivity have been identified (see \cite{FrenchPavlidis} for results  based on connectivity data in the
 rat brain).
It would be interesting to  investigate to which extent these correlations persist across the loops that we identified in the mouse brain. 
 Moreover, spatial gene expression patterns have been related to cortico-striatal functional networks \cite{cortexStriatum}. 
  Loops that are as  persistent as the cortico-striatal ones yield natural  candidates
  for places where to look for such  genetic signatures of functional networks. Moreover, the brain-wide coverage of the gene-expression
 atlas \cite{AllenGenome,AllenAtlasMol} allows to estimate spatial densities of cell types
  known by their transcriptional activity \cite{integrationPNAS,toolbox,methods}, which yields neuroanatomical 
 predictions relevant to brain disorders \cite{coExpression,cellTypeNext}. The present work could lead to 
 insights into the correlation structure between the local density of loops and the spatial density of cell types. Another 
  direction of research is the conservation of  patterns across species \cite{complexNetwork,largeScale,canonical}, even though
  gene-expression and connectivity data do not have the same brain coverage as in the mouse atlas.\\
 
We have worked out the persistent loops of the wiring diagram of the brain based on 
 ipsilateral projections only, leaving out the issue of contralateral proections. Contralateral 
 projections have been mapped (see \cite{AllenConnectome}, right-hand side of Fig. 4A), but they do not allow
 to work out closed loops because the missing projections would involve injecting tracers into the 
  left hemisphere, and mapping the contralateral projections again. However, the known contralateral connection 
 strengths have been found to be significantly weaker than ipsilateral ones. One can therefore conjecture than
  loops going through brain regions in different hemispheres do not dramatically change the global features
  of the wiring diagram of the brain.

There are surprisingly few cycles confined to only one of of the major brain regions (or big regions),
 but most of them develop before the brain becomes fully connected. 
 Moreover, medulla is by far the region with the most confined loops.  

%
%

\section{Acknowledgments}
  This work is supported by the Research Center for Precision Medicine, HT-URC, {\hbox{Xi’an Jiaotong-Liverpool University}}, Suzhou, China.


%
%
%
\afterpage{\clearpage}

\end{document}